# Comment on 3D Penrose Tiling of the Icosahedral Quasicrystalline Phase

L. J. Swartzendruber[1] and L. H. Bennett[2*],

1. NIST (Retired)
2. Institute for Magnetics Research, The George Washington University, Ashburn, USA
*Corresponding author: lbennett@gwu.edu

*Abstract*

*A recent arXiv contribution makes the case that the quasicrystalline phase is due to twinning, albeit with a different twinning scheme than that presented by Pauling. It appears to us that this approach, though novel, has no validity.*

The 2011 Nobel Prize in Chemistry was awarded to Dan Shechtman for his discovery [1] in 1982 of an icosahedral phase in splat-cooled $Al_6Mn$, which created the new field of quasiperiodic crystals. This atomic arrangement was completely unexpected and had, until then, been considered to be impossible to form under the laws of crystallography. It has icosahedral point symmetry with six five-fold axes, and decagonal point symmetry with one 10-fold axis. Shechtman melted Al and Mn and rapidly cooled the mixture. Because the sample size was very small, he used electron microscopy and found an icosahedral diffraction pattern of sharp spots. The symmetry of the pattern implies quasiperiodicity. Paul Steinhardt, and Dov Levine, [2] came up with a possible explanation. Shechtman's quasicrystals were imperfect, however, and the data could not rule out other explanations, like "twinning," in which two periodic crystals are fused at an angle. There were considerable objections to the conclusion that a new crystalline phase was involved. "There is no such thing as quasicrystals," said Linus C. Pauling, the winner of two Nobel Prizes, "only quasi scientists." [3] He presented detailed "proof" [4] of the twinning behavior he proposed, which involved a large unit cell of at least 1,000 atoms, with multiple twinning to mimic icosahedral symmetry..

Subsequently, An-Pang Tsai, at Tohoku University, discovered [5] a new alloy made of iron, copper and aluminum — "the first example of a true, bona fide quasicrystal," said Dr. Steinhardt [6]. A Mössbauer study [7] produced spectra consistent with the quasicrystalline model, but inconsistent with the transition metals at the center of iosahedra. An NMR experiment [8] comparing icosahedral Al-Mn with the crystalline G-phase $Al_{12}Mn$ provided very strong evidence that the large cubic twinned crystal model of Pauling was not responsible for the icosahedral phase [7]. In the G-phase, each Mn atom is surrounded by an icosahedron of Al atoms which is exactly as in Pauling's proposed structure. Pauling maintained his objection, but most crystallographers eventually concluded that the quasicrystal was a "crystal", and not an artifact of twinning. The International Union of Crystallography has now acknowledged that the quasicrystal is, in fact, a crystal! Even Pauling was almost convinced, writing [3] 'Professor Shechtman, may I propose to you to write the joint Shechtman-Pauling paper on quasi-periodic materials? And you will be first,' he says. Unfortunately Pauling's failing health prevented such a collaboration from taking place

The five and ten fold rotational symmetries observed by Shechtman in rapidly solidified $Al_6Mn$ were forbidden by the accepted rules of crystallography because they could not occur in crystals with periodic unit cells. As shown by Penrose [9] forbidden five fold symmetries were obtainable in an aperiodic 2D tiling using prolate and oblate golden rhombuses, one with an acute angles of 2π/5, the other with an acute angles of 2π/10. In analogy with the 2D Penrose tiling, an aperiodic 3D tiling was proposed using two golden rhombohedra, a prolate with solid angles of π/5 steradians at the ends of the one long diagonal, and an oblate with solid angles of 7π/5 steradians at the ends of the one short diagonal.  The now generally accepted approach includes such tiling obtained by making 3D projections from 6D periodic structures, which was used recently, for example, by Nagao et al. [10] in analyzing *in situ* observations of the growth of an AlNiCo quasicrystal using high resolution transmission electron microscopy.

But there is a new twinning explanation raised at this late date.  Prodan *et al.* [11] claim that Pauling was correct explaining quasicrystals by multiple twinning, but that he should not have based the twinning on the overall structure, but instead have applied twinning at the level of the smallest golden rhombohedra.  They carried out a 3D Penrose tiling using the prolate and the oblate primitive golden rhombohedra and their diffraction patterns *separately* for all possible twinned orientations.  Then they combined these into multiply twinned star polyhedra constructed by overlapping individual contributions.  They get a calculated diffraction pattern which is very close to that seen in icosahedral quasicrystals.  Their pattern becomes much closer by introducing phason distortions The now generally accepted approach also includes tiling with two golden rhombohedra by making 3D projections from 6D periodic structures, so the Prodan *et al.* approach, though interesting, provides no support that twinning has any validity in a correct explanation.

Although this is an interesting proposal major objections are apparent.  First of all, even if their proposed structure actually existed, their multiple "twinning at the unit cell level" is an extension of classical crystallography and not a confirmation of Pauling's crystalline twinning idea.  Furthermore, if there is a specific long range order of the constituent golden rhombohedra, this order should be detectable.  To our knowledge, no such order has been observed by x-ray diffraction or by other means.  In addition, the proposed structure does not explain the observation of five-fold faces observed in quasicrystals more stable than $Al_6Mn$, such as HoMgZn [12] which grows to have macroscopic faces that are true regular pentagons.

It is interesting to point out that, starting with a nonconvex rhombic hexacontahedron [13] (which consists of twenty golden prolate rhombohedra and is central to the approach of Prodan *et al.*) all space can be filled aperiodically in a way that preserves the six fivefold rotational axes of the original hexacontahedron using only prolate rhombohedra.